\input harvmac.tex

\input epsf.tex

\def\figin{\epsfcheck\figin}\def\figins{\epsfcheck\figins}
\def\epsfcheck{\ifx\epsfbox\UnDeFiNeD
\message{(NO epsf.tex, FIGURES WILL BE IGNORED)}
\gdef\figin##1{\vskip2in}\gdef\figins##1{\hskip.5in}% blank space instead
\else\message{(FIGURES WILL BE INCLUDED)}%
\gdef\figin##1{##1}\gdef\figins##1{##1}\fi}
\def\DefWarn#1{}
\def\figinsert{\goodbreak\midinsert}
\def\ifig#1#2#3{\DefWarn#1\xdef#1{fig.~\the\figno}
\writedef{#1\leftbracket fig.\noexpand~\the\figno}%
\figinsert\figin{\centerline{#3}}\medskip\centerline{\vbox{\baselineskip12pt
\advance\hsize by -1truein\noindent\footnotefont{\bf Fig.~\the\figno:} #2}}
\bigskip\endinsert\global\advance\figno by1}

\font\sf = cmss10
%%%%%%%%%%%%%%%%%%%%%%%%%%%%%%%%%%%%%%%%%%%%%%%%%%%%%%%%%%%%%%%%%%%%%%%%%%%%%

%\draftmode

%%%%%%%%%%%%%%%%%%%%%%%%%%%%%%%%%%%%%%%%%%%%%%%%%%%%%%%%%%%%%%%%%%

{ \Title{\vbox{\baselineskip12pt \hbox{hep-th/0508058 }
\vbox{\baselineskip12pt \hbox{UB-ECM-PF-05/20 }}
%{\vbox{\baselineskip12pt \hbox{ecm } }}
}}
 {\vbox{
{\centerline { A 1+1 field theory spectrum from M theory}
%{\centerline {from M theory} }
}}}}

\bigskip
\centerline{ 
Maria Jose Rodriguez$^a$\footnote{$^{\ast}$}{majo@ecm.ub.es}
 and Pere Talavera$^b$\footnote{$^{\star}$}{pere.talavera@upc.edu} }
\bigskip~

\centerline{$^a$
Departament d'Estructura i Constituents de la Mat\`eria,}
\centerline{
Universitat de Barcelona,
Diagonal 647, E-08028 Barcelona, Spain}

\medskip

\centerline{$^b$ Departament de F{\'\i}sica i Enginyeria Nuclear,}
\centerline{ Universitat Polit\`ecnica de Catalunya, Jordi Girona
1--3, E-08034 Barcelona, Spain }

\vskip .3in

\baselineskip12pt
\vfill

The spectrum of a 1+1 dimensional field 
theory with dynamical quarks is constructed. We focus in testing the possible brane embeddings that can support fundamental matter. The requirement on the wave function normalisation  and
the dependence on the quark mass of the quark condensate
allow to discard most of the embeddings. We pay attention to some more
general considerations comparing the behaviour of the non-compact theory at
different dimensions. In particular 
we explored the possibility that the AdS/CFT duality ``formalism''  introduce a scale breaking parameter at $(1+1)d$ allowing the existence of {\sl classical} glueballs and its possible relation with point-like string configurations. 
The screening effects and the appearance of a possible phase transition
is also discussed.

\vfill

\Date{August  2005} \eject \baselineskip14pt

%\listtoc
%\writetoc

\newsec{Motivation and conclusions}

%\ZhitnitskyUM
\lref\ZhitnitskyUM{
  A.~R.~Zhitnitsky,
  ``On Chiral Symmetry Breaking In QCD In Two-Dimensions (N(C) $\to$
  Infinity),''
  Phys.\ Lett.\ B {\bf 165}, 405 (1985)
  [Sov.\ J.\ Nucl.\ Phys.\  {\bf 43}, 999.1986\ YAFIA,43,1553 (1986\ YAFIA,43,1553-1563.1986)].
  %%CITATION = PHLTA,B165,405;%%
}

%\HorowitzBJ
\lref\HorowitzBJ{
  G.~T.~Horowitz and H.~Ooguri,
  ``Spectrum of large N gauge theory from supergravity,''
  Phys.\ Rev.\ Lett.\  {\bf 80}, 4116 (1998)
  [arXiv:hep-th/9802116].
  %%CITATION = HEP-TH 9802116;%%
}

%\'tHooftHX
\lref\tHooftHX{
  G.~'t Hooft,
``A Two-Dimensional Model For Mesons,''
  Nucl.\ Phys.\ B {\bf 75}, 461 (1974).
  %%CITATION = NUPHA,B75,461;%%
}

%\ColemanCI
\lref\ColemanCI{
  S.~R.~Coleman,
``There Are No Goldstone Bosons In Two-Dimensions,''
  Commun.\ Math.\ Phys.\  {\bf 31}, 259 (1973).
  %%CITATION = CMPHA,31,259;%%
}

%\WittenZW
\lref\WittenZW{
  E.~Witten,
``Anti-de Sitter space, thermal phase transition, and confinement in  gauge
 theories,''
  Adv.\ Theor.\ Math.\ Phys.\  {\bf 2}, 505 (1998)
  [arXiv:hep-th/9803131].
  %%CITATION = HEP-TH 9803131;%%
}

%\GrinsteinXK
\lref\GrinsteinXK{
  B.~Grinstein and R.~F.~Lebed,
  ``Explicit quark-hadron duality in heavy-light meson weak decays in the  't
  Hooft model,''
  Phys.\ Rev.\ D {\bf 57}, 1366 (1998)
  [arXiv:hep-ph/9708396].
  %%CITATION = HEP-PH 9708396;%%
}

%\DalleyYY
\lref\DalleyYY{
  S.~Dalley and I.~R.~Klebanov,
``String spectrum of (1+1)-dimensional large N QCD with adjoint matter,''
  Phys.\ Rev.\ D {\bf 47}, 2517 (1993)
  [arXiv:hep-th/9209049].
  %%CITATION = HEP-TH 9209049;%%
}

%\BrandhuberER
\lref\BrandhuberER{
  A.~Brandhuber, N.~Itzhaki, J.~Sonnenschein and S.~Yankielowicz,
  ``Wilson loops, confinement, and phase transitions in large N gauge  theories
  from supergravity,''
  JHEP {\bf 9806}, 001 (1998)
  [arXiv:hep-th/9803263].
  %%CITATION = HEP-TH 9803263;%%
}

%\ParedesIS
\lref\ParedesIS{
  A.~Paredes and P.~Talavera,
  ``Multiflavour excited mesons from the fifth dimension,''
  Nucl.\ Phys.\ B {\bf 713}, 438 (2005)
  [arXiv:hep-th/0412260].
  %%CITATION = HEP-TH 0412260;%%
}

%\DeserWQ
\lref\DeserWQ{
  S.~Deser,
  ``Absence Of Static Solutions In Source - Free Yang-Mills Theory,''
  Phys.\ Lett.\ B {\bf 64}, 463 (1976).
  %%CITATION = PHLTA,B64,463;%%
}

%\ColemanHD
\lref\ColemanHD{
  S.~R.~Coleman,
  ``There Are No Classical Glueballs,''
  Commun.\ Math.\ Phys.\  {\bf 55}, 113 (1977).
  %%CITATION = CMPHA,55,113;%%
}

%\PagelsCK
\lref\PagelsCK{
H.~Pagels,
``Absence Of Periodic Solutions To Scale Invariant Classical Field Theories,''
Phys.\ Lett.\ B {\bf 68}, 466 (1977).
%%CITATION = PHLTA,B68,466;%%
}

%\BabingtonVM
\lref\BabingtonVM{
  J.~Babington, J.~Erdmenger, N.~J.~Evans, Z.~Guralnik and I.~Kirsch,
  ``Chiral symmetry breaking and pions in non-supersymmetric gauge /  gravity
  duals,''
  Phys.\ Rev.\ D {\bf 69}, 066007 (2004)
  [arXiv:hep-th/0306018].
  %%CITATION = HEP-TH 0306018;%%
}

%\CallanPS
\lref\CallanPS{
  C.~G.~.~Callan, N.~Coote and D.~J.~Gross,
  ``Two-Dimensional Yang-Mills Theory: A Model Of Quark Confinement,''
  Phys.\ Rev.\ D {\bf 13}, 1649 (1976).
  %%CITATION = PHRVA,D13,1649;%%
}

%\GrossGK
\lref\GrossGK{
  D.~J.~Gross and H.~Ooguri,
  ``Aspects of large N gauge theory dynamics as seen by string theory,''
  Phys.\ Rev.\ D {\bf 58}, 106002 (1998)
  [arXiv:hep-th/9805129].
  %%CITATION = HEP-TH 9805129;%%
}

%\PolchinskiUF
\lref\PolchinskiUF{
  J.~Polchinski and M.~J.~Strassler,
  ``The string dual of a confining four-dimensional gauge theory,''
  arXiv:hep-th/0003136.
  %%CITATION = HEP-TH 0003136;%%
}

%\AharonyUP
\lref\AharonyUP{
  O.~Aharony,
  ``The non-AdS/non-CFT correspondence, or three different paths to QCD,''
  arXiv:hep-th/0212193.
  %%CITATION = HEP-TH 0212193;%%
}
There has been a lot of effort in trying to understand the low-energy
dynamics of non-supesymmetric
asymptotically free gauge theories from the supergravity
duals ( see for instance \refs{\GrossGK, \WittenZW, \PolchinskiUF,\AharonyUP}).
Most of these studies are focous
at dimensions $d=4,3$ where confinement can only be understood
as a dynamical feature while the case
of dimension $d=2$ is relegated to oblivion, probably because confinement
is automatically incorporated in the theory even at perturbative level. The most salient point is precisely that the theory  is fully resoluble in a suitable limit.

Field theory models at $d=2$
while retaining some features of the four dimensional
QCD theory as quark confinement and chiral symmetry breaking, still differs
substantially in many crucial aspects from it: {\sl i)} there are no
dynamical gluons and hence strings are only build from matter quanta alone.
{\sl ii)} There is no chromomagnetic field, a key point to understand
confinement in 3+1 dimensions and {\sl iii)}
it neither contains spin.

We want to elaborate here on the physical spectra, vectors, massive scalars
and possible glueballs, one can find
when dealing with one of these non-supersymmetric confining backgrounds
in $d=2$ \WittenZW. We shall argue, that parallel to the field theory expectations,  the supergravity model properties are as follows: there is spontaneous symmetry breaking, there are massless particles, the physical vectors acquires a mass.
There are, obviously, still many lacking desired features but is encouraging to obtain some of the expected field theory results emerging from  brane configurations.

In order to illustrate the field theory side we are after for, we review it briefly.

\subsec{Field theory set up}

The field theory parallel we want mainly to study is the 1+1 dimensions
SU(N$_c$) gauge theory coupled to matter field in the
fundamental representation, \refs{\tHooftHX,\CallanPS} namely the 't Hooft
model. Although our initial setup will be related to the
SU(N$_c$) gauge field coupled to static fermions in the adjoint representation
\DalleyYY .

The main difference between both settings are found in their interpretation as
string models:
while the 't Hooft model can be thought of as an open string model
with a single rising Regge trajectory spectrum, the coupling to adjoint matter
produces a kind of closed string with multiple Regge trajectories.

In the case of SU(N$_c$) YM coupled to fundamental matter, our main
purpose, the Minkowsky
space action is taken to be
\eqn\qcddos{
S = \int d^2 x {\rm Tr}\left[ \bar{q}\left(i \gamma^\mu D_\mu-m_q\right)
q -{1\over 2 g_{{\rm
strong}}^2} F^{\mu\nu} F_{\mu\nu}\right]\,,
}
where $F_{\mu\nu}=\partial_\mu A_\nu- \partial_\nu A_\mu+
i[A_\mu,A_\nu]$ and the covariant derivative is $D_\mu \Phi= \partial_\mu
\Phi+ i[A_\mu,\Phi]\,.$
In this case the
quark-antiquark sector admits an infinite tower of confined color-singlets.

The reason for this solubility lies in the very defining features of the model.
The large N$_c$ limit eliminates all the sea quark contributions together
with the non-planar gluon diagrams. On the other hand the
fact of being in an axial gauge, $A_+=0$, and being the action independent
of $x_-$ derivatives allows to disentangle the gluon
self-coupling by gauging away $A_-\,.$ The only remaining Feynman diagrams
in the 2-PI Green function are the rain-bow and ladder type, whose
Schwinger-Dyson equation, giving the meson spectrum, can be solved
in principle numerically.

\medskip

The content of the paper is: we first start reviewing the construction of the non-supersymmetric background in a very synthetic way, it mainly should introduce the notation we shall follow and all the possible relevant embeddings. In sec. 3 a possible physical embedding providing matter in the fundamental is introduced. We check that  from the initial two possibilities only one is feasible due to the normalisation at infinity. For the remaining solution we look for the relation between the chiral condensate and the quark mass. After comparison with the field theory and lattice calculations we find that the obtained results do not follow their trends.  In accordance we turn to the evaluation of other embeddings in sec. 4. Once more the behaviour of the wave function at infinite imposes serious restrictions on the allowed solutions. For the remaining solutions we find the massive scalar and vector mesons spectra.

Thenceforth
the subjects we treat has no special relation of been
 treating a 1+1 dimensional theory, 
but on the fact of having added fundamental matter.
In sec. 5, following field theory arguments, we argue on the possible existence of classical glueballs. We present both, numerical and analytical evidence that the formalism can not account to their existence.  Although  we dot not obtain glueball modes we check in sec. 6 the existence of oscillating modes on a ``cigar'' type configuration. This configuration was suggested to be related, at least in the four dimensional case, to the supergravity glueball spectra. The   existence of the screening effect, hence the existence of at least two different phases, is  discussed in sec. 7.

\newsec{Non-extremal Dp-brane model }

We shall introduce the main notation for the different quantities used in the text.
The starting point is the metric and the dilaton field
\eqn\metricex{ds^2 = h^{-1/2} dx_\parallel^2
+h^{1/2}\left(dU^2 + U^2 d\Omega_{8-p}^2\right)\,,\quad e^{\phi}=\left(2\pi\right)^{2-p} g_{\rm YM}^2 h^{(3-p)/4}\,.
}
The transverse space to the brane has dimension $9-p\,,$
and the wrap factor is given by
\eqn\wrap{h^{-1}={U^{7-p}\over \left(g_{\rm YM} \sqrt{d_p N_c}\right)^2}\,.
}
 In the remainder it will probe
useful to define $R_p^{7-p}=\left(g_{\rm YM} \sqrt{d_p N_c}\right)^2\,.$

The non-extremal metric is obtained by imposing anti-periodic boundary conditions on the adjoint fermions, so that they become massive  \WittenZW\
\eqn\metricnonex{ds^2 = h^{-1/2} \left(dy_\parallel^2+f(U) d\theta_2^2\right)
+h^{1/2}\left({dU^2\over f(U)} + U^2 d\Omega_{8-p}^2\right)\,,
}
where the function $f$
\eqn\ffunction{f(U) = 1-\left(U_h\over U\right)^{7-p}\,,}
contains an IR scale, $U_h\,,$ that breaks
conformal invariance.
Notice that at very high energy, $U\gg U_h$,
\metricnonex\ reduces to \metricex .
To avoid the conical singularity the period of the compact $\theta_2$
 variable is chosen to be $\delta \theta_2 = 4 \pi/(7-p) (R_p/U_h)^{(7-p)/2}
U_h\,.$ This compactification has a mass scale associated
\eqn\kkmas{
m_{KK}= {2\pi\over \delta\theta_2}={1\over 2 U_h} (7-p)
\left({U_h \over R_p}\right)^{(7-p)/2} \,.
}
{}For energy scales lower than $m_{KK}$ the theory is effectively $p$-dimensional.
The transverse part of the Dp-brane can be parametrised by a set of coordinates
$\vec{z}=\left(z^1,\dots,z^{9-p}\right)$ suitable to write it
as conformally flat
\eqn\metricnonextwo{ds^2 = h^{-1/2} \left(dy_\parallel^2+f(U) d\theta_2^2\right)
+K(\rho) d\vec{z}\cdot d\vec{z}\,,
}
where $\rho^2 = \vec{z}\cdot \vec{z}$~ and $d\vec{z}\cdot d\vec{z} = d\rho^2
+\rho^2 d\Omega_{8-p}^2\,.$
In order to obtain \metricnonextwo\ one identifies
\eqn\relchange{K(\rho) = h^{1/2}\left({U\over \rho}\right)^2\,,
}
and perform the change of variables
$
\partial_\rho U=U \sqrt{f(U) }/\rho\,.
$
When inserting \ffunction\ in the previous expression one gets
\eqn\ufrho{U(\rho)= \left[ {1\over 2}
\left({U_h\rho\over A}\right)^{{7-p\over 2}}
+ {1\over 2}\left({U_hA\over \rho}\right)^{{7-p\over 2}} \right]^{2\over 7-p}\,,
}
%\KruczenskiUQ
\lref\KruczenskiUQ{
  M.~Kruczenski, D.~Mateos, R.~C.~Myers and D.~J.~Winters,
``Towards a holographic dual of large-N(c) QCD,''
  JHEP {\bf 0405}, 041 (2004)
  [arXiv:hep-th/0311270].
  %%CITATION = HEP-TH 0311270;%%
}
with $A$ an integration constant. In order to conform the notation in
\KruczenskiUQ\ we set $\left(U_h/A\right)^{{7-p\over 2}}=2\,,$ from where
\eqn\ufrhofinal{U(\rho)= \rho \left[ 1
+ {1\over 4}\left({U_h\over \rho}\right)^{7-p} \right]^{{2\over 7-p}}\,.
}

\bigskip
%\KarchSH
\lref\KarchSH{
  A.~Karch and E.~Katz,
``Adding flavor to AdS/CFT,''
  JHEP {\bf 0206}, 043 (2002)
  [arXiv:hep-th/0205236].
  %%CITATION = HEP-TH 0205236;%%
}

To incorporate matter in the fundamental representation one embeds
a probe-brane wrapping trivial circles on the transverse space of
\metricnonextwo\ \KarchSH.
Bearing that in mind we shall depict the transverse space in a way that respect
the symmetries of the would be embedded-brane. As previously mentioned
we are interested in study the parallel of QCD$_{1+1}$. The responsible
of the background will be a stack of D2-branes and
the possible embeddings are given by the arrows

\medskip
$$\vbox{
\halign{ # & \quad # & $\, $# & $\, $#& $\, $#& $\,$ #& $\,$ #
& $\,$ #& $\,$ #& $\,$  # &$\,$  #  \cr
 D2: & 0 & 1 & 2 & $\_$ & $\_$ & $\_$ & $\_$ & $\_$ & $\_$ & $\_ \,$ \cr
 D4: & 0 & 1 & $\_$ & 3 & 4 & 5 & $\_$ & $\_$ & $\_$ & $\_ \,$ \cr
 D6: & 0 & 1 & $\_$ & 3 & 4 & 5 & 6 & 7 & $\_$ & $\_ \,$ \cr
 D8: & 0 & 1 & $\_$ & 3 & 4 & 5 & 6 & 7 & 8 & 9 \cr
}}$$
\medskip

Attending to the symmetries of the possible embedded Dm-probe brane
 the background field metric is written as
\eqn\metricdtwo{
ds^2_{{\rm p}=2} = h^{-1/2}
\left( ds^2\left({\sf E}^{(1,1)}\right)+f(U) d\theta_2^2\right) + K(\rho) \left(
d\Omega_{m-1}^2 + d\Omega_{8-m}^2 \right)\,,
}
Being from now on $\lambda$~and~$r$~the radii of the transverse n-spheres to the
Dp-brane, with $\rho^2=\lambda^2 +r^2\,.$

\newsec{Chiral symmetry breaking}

The embedding of $N_f$  Dm-probe brane in the ambient space of $N_c$ Dp-brane
($p\le m$) can be approximated by the DBI action
\eqn\dbi{
S^{(p)}_{\rm Dm}
= -{1\over (2\pi)^m g_s \ell_s^{m+1}} \int d^{m+1}\sigma e^{-\phi}
\sqrt{-{\rm det} g}\,,
}
if $N_f$ is held fixed and $N_c \gg N_f$. In \dbi\
$g$ refers to the pullback metric. For later purposes
the Dm-brane tension is defined as
$T_{\rm Dm} = \left((2\pi)^m g_s \ell_s^{m+1}\right)^{-1}\,,$ and its
world-volume coordinates will be parametrised by
$\sigma^{0,\ldots\,,m}=x^{0,\ldots\,,p-1},z^{p,\ldots \,, m}\,.$

In the sequel, and to perform the embedding,
 we shall deal with an ansatz concerning the position
of the Dm-brane in the space spanned for the latest n-sphere in
\metricdtwo\ . This n-sphere will be parametrised
by the radii $r$~and the angles $\phi^a$ $(a=0,\dots\,,n-2)\,.$ Then the
ansatz is chosen to be \BabingtonVM
\eqn\ansatzone{ r(\lambda)\,,\quad \phi^a={\rm fixed}\,,\quad\tau=
{\rm fixed}\,,
}
where the latest condition will be relaxed afterwards.

\medskip

\subsec{Embedding on D2-branes}

The possible embeddings on a  D2-brane bulk geometry were sketched previously
and their action read
\eqn\embededmetddos{
S_{\rm Dm}= -T_{\rm Dm} \int d^{m+1}\sigma
\left({R_2\over \rho}\right)^{5(m-4)/4} \left(1+{U_h^5\over 4 \rho^5}
\right)^{(16-m)/10}\lambda^{m-2} \sqrt{1+\dot{r}^2} \sqrt{h_{\Omega_{m-2}}}
\,,
}
with $h_{\Omega_{m-2}} $ the determinant on a unit (m+1) dimensional sphere.
\smallskip
The classical
equation of motion for the transverse, $r(\lambda)\,,$ mode is
$$
{d\over d\lambda}\left[{\lambda^{m-2}\over\rho^{5(m-4)/4}}
\left(1+{1\over 4}
\left({U_h\over\rho}\right)^5\right)^{(16-m)/10}
{\dot{r}\over {\sqrt{1+\dot{r}^2}}}\right]=
-{1\over 16 \rho^2} U_h^5
$$
\eqn\eommfor{ {\lambda^{m-2}\over\rho^{5m/4}} \left(1+{1\over 4}
\left({U_h\over\rho}\right)^5\right)^{(6-m)/10} r\,
\sqrt{1+\dot{r}^2} \left(3(4+m) + 20 (m-4)\left({\rho\over
U_h}\right)^5\right)\,. } The BPS case is recovered by setting
$U_h=0\,,$ then a particular solution to \eommfor\ is $r(\lambda)=
{\rm constant}$ reflecting that no force acts on the Dm-brane. To
analyze further \eommfor\ is worth to rescale it as $\lambda\to U_h
\lambda \,, r \to U_h r \,, \rho\to U_h\rho\,. $ The differential
equation is non-linear and it was not possible to find an analytic
solution to \eommfor\ therefore we analyzed it asymptotically.
More in concrete, we search for solutions in the asymptotic region
with finite distance between the D2 and the Dm branes. That is
$\lambda\to\infty\,,~ r(\lambda)\to r_\infty\,. $ This in turn
implies $\dot{r}\to 0$~ and $\rho\sim\lambda\,$ which can be thought
as the linearisation of the equation. 
With the above behaviour \eommfor\ becomes
\eqn\eommforasy{ {d\over d\lambda} \left[\lambda^{3-m/4}\,
\dot{r}\right]= -{1\over 16} r(\lambda) {1 \over \lambda^{m/4+4}}
\left( 3(4+m) + 20 (m-4) \lambda^5\right) \,. } 
Notice that 
$\lambda \to \infty\,$ appears as an irregular (regular)
singular point in the D6-brane (D4-brane) 
embedding signaling some kind of illness in its solutions.
{}For the two
possible non-trivial embbedings one obtains 
\eqn\eommforasyonly{
{\rm D4}:\quad {d\over d\lambda} \left[\lambda^2 \dot{r}\right]= -{3\over
2} {r(\lambda)\over \lambda^5} \,, \qquad {\rm D6}:\quad {d\over d\lambda}
\left[\lambda^{3/2} \dot{r}\right]= -{5\over 2} {r(\lambda)\over
\lambda^{1/2}} \,, } while for the former one can find a
normalisable solution for the radial coordinate \eqn\soleommforasy{
r(\lambda) = A {1\over\sqrt{\lambda}} J_{-{1\over 5}}\left(
{\sqrt{6}\over 5} {1\over \lambda^{5/2}}\right) + B
{1\over\sqrt{\lambda}} J_{{1\over 5}}\left( {\sqrt{6}\over 5}
{1\over \lambda^{5/2}}\right)\,, } the latter embedding presents an
asymptotic oscillatory behaviour thus lacking any physical
interpretation. We shall focous on the D4 case in the remainder.

The D4 embedding case reduces to,
for arbitrary large $\lambda$ values,
\eqn\approxsoleommforasy{
r(\lambda) = r_\infty + {c\over \lambda}\,.
}
The physical interpretation of the coefficients $r_\infty$~and $c$ is the same
as in the QCD$_{3+1}$ case \KruczenskiUQ\ : they are related to the quark mass
and the chiral condensate respectively
\eqn\physc{
m_q = {U_h\over 2\pi} r_\infty\,,\quad {\delta {\cal E}\over \delta m_q}=\langle \bar{q}q\rangle=
-8\pi^2 T U_h^2 c\,.
}

In order to gain more intuition on the solution, we inspect the region
$r_\infty\gg$. This must be a small perturbation to the BPS state. Inserting
$r(\lambda) = r_\infty+\delta r(\lambda)$ in \eommfor\
leads, at leading order in $\delta r$,
\eqn\leadingdquatre{
{d\over d\lambda}\left(\lambda^2 \delta\dot{r}\right) \approx
-{3\over 2} r_\infty {\lambda^2\over\left(r_\infty^2+\lambda^2\right)^{7/2}
}\,.
}
Integrating with the boundary conditions: {\it i)} $\dot{r}\vert_{\lambda=0}=0$
and {\it ii)} $\delta r\vert_{\lambda\to\infty}=0$ gives the final answer
\eqn\finaladquatre{r(\lambda)  \approx r_\infty+
{3r_\infty^2 + 2 \lambda^2\over 10 r_\infty^3
\left(r_\infty^2+\lambda^2\right)^{3/2}}\,.
}
{}From the asymptotic result we can match the coefficients of the previous
relation with those of \approxsoleommforasy\ getting $c(r_\infty) \sim
1/(5r_\infty^3)$, i.e. the chiral condensate scales as $1/m_q^3\,,$ for large quark masses.
We shall comment on that behaviour latter on in connection to the field theory
expectations.
Notice that \finaladquatre\ has  negative (positive)
derivative for increasing positive (negative)
values of $\lambda$, expecting then, a kind of bump or repulsion between the
D4 and the D2-branes near the origin.

\ifig\ejdos{In the l.h.s. panel there is plotted
the D4 profile in the 8-9 plane. The D2-brane is the dashed line. In the r.h.s. we plotted the scaling of the quark condensate with respect to the quark mass.
}{
\epsfxsize 2.7 in\epsfbox{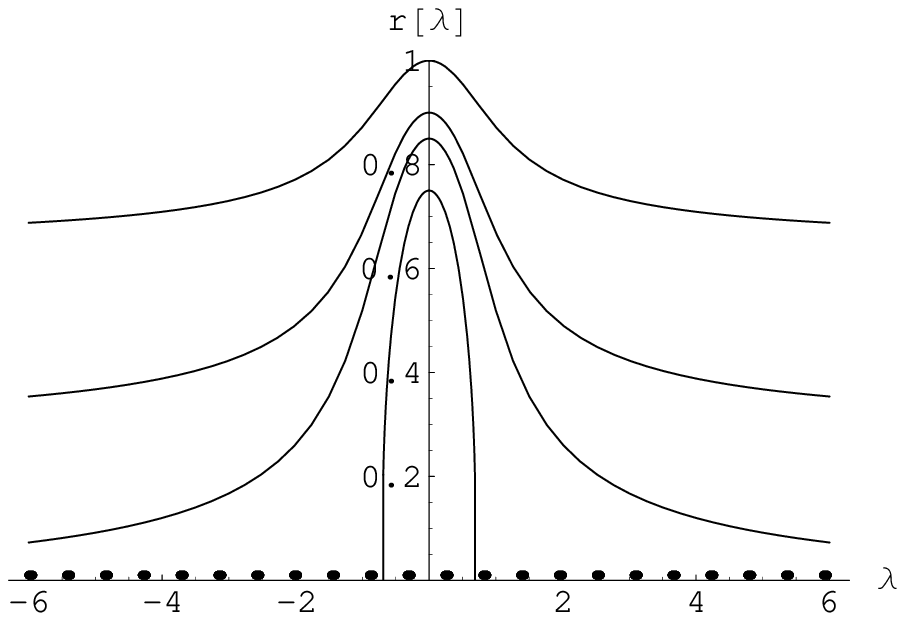}
\epsfxsize 2.7 in\epsfbox{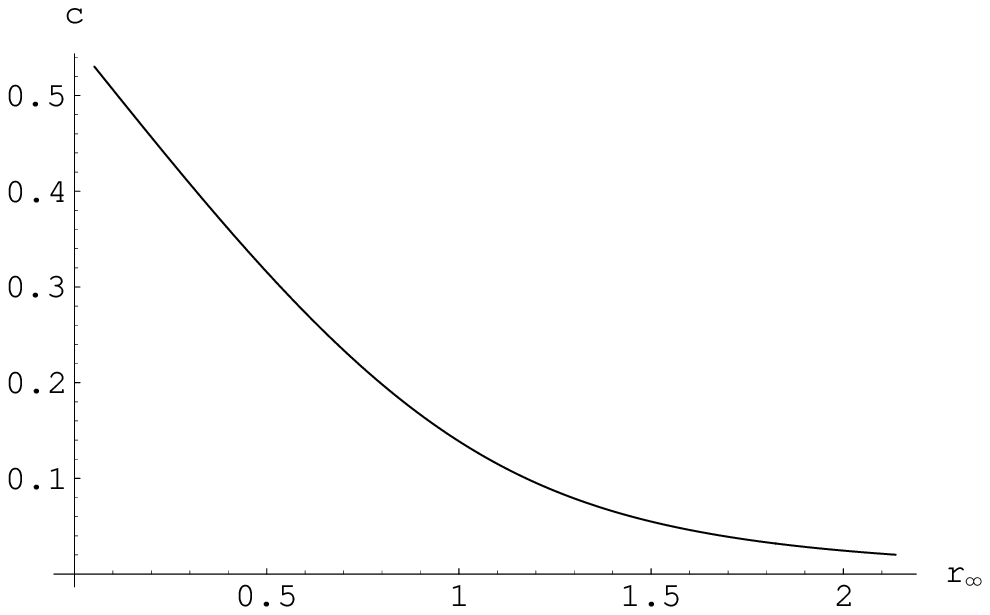}}

\smallskip

In \ejdos\ l.h.s.
we depicted the profile of the D4 probe-brane in the $8-9$ transverse
plane. The D2-brane is located on the x-axis.  In the r.h.s. we have evaluated numerically
the coefficients $c$ and $r_\infty$ coming from \eommfor. The numerical results
corroborates all the previous analytical findings.

%\bigskip
%\noindent {\it 3.1.1. Spheric approximation}
%\medskip

\bigskip

%\subsec{Field theory interpretation}

We can compare without further ado the previous findings with the corresponding
field theory. At high energy or in processes involving heavy-quarks
is well known that the use of summe-rules can provide the desired
link between the hadron observables in terms of the fundamental degrees of freedom.
{}For heavy-quarks, the quark condensate behaves as $\langle \overline{q} q\rangle \sim
\langle FF\rangle /m_q^r$ been $r$ a c-number fixed on dimensional grounds. The dimensional counting derived from \qcddos~ leads to: $[q]=E^{1/2}$ and $[F]=E$ hence it follows that $r=1\,,$ {\sl i.e.} $\langle \overline {q} q\rangle \sim 1/m_q\,.$ The previous expressions does not match with its purported supergravity dual.

\newsec{$\overline{{\rm Dm}}$-D2-Dm system}

%\SakaiCN
\lref\SakaiCN{
  T.~Sakai and S.~Sugimoto,
``Low energy hadron physics in holographic QCD,''
  arXiv:hep-th/0412141.
  %%CITATION = HEP-TH 0412141;%%
}
It is clear that \ansatzone~ is not able to reproduce the proper quark condensate in $1+1$ consequently we shall turn now to another ansatz involving the compact direction
\KruczenskiUQ . In the 4 dimensional case this was throughly
explored in \SakaiCN.
The embedding for the D4, D6 and D8 (generically denoted Dm) in the bulk of
the D2 with $U(\theta_2)$ gives an action
%\eqn\embetau{ds^2_{\rm Dm} = h^{-1/2} dy_\parallel^2 + \left(
%h^{-1/2} f(U) + {h^{1/2}\over f(U)} \left(\partial_{\theta_2}U\right)^2
%\right) d\theta_2^2 + h^{1/2} U^2 d\Omega_{m-2}^2\,.
%}
%The action \dbi\ reads with \embetau\
\eqn\actiontau{
S_{\rm Dm}= -{T_{\rm Dm}\over g_{\rm YM}^2} \int d^{m+1}\sigma \left( f(U) +
{h\over f(U)} \left(\partial_{\theta_2}U\right)^2 \right)^{1/2} U^{m-2}
h^{(m-6)/4} \sqrt{h_{\Omega_{m-2}}}\,,
}
that does not depend explicitly on $\theta_2\,.$
This last fact allows to obtain a first integral of motion
%\eqn\hamil{
%H(U)= - {U^{m-2}f(U) h^{(m-6)/4}\over \sqrt{f(U)+{h\over f(U)}
%\left(\partial_{\theta_2}U\right)^2 }} \sqrt{h_{\Omega_{m-2}}}\,.
%}
\eqn\taut{
\theta_2(U)= \sqrt{f(U_0) U_0^{11-m/2}} \int_{U_0}^U {dU\over
h^{-1/2} f(U) \sqrt{f(U) U^{11-m/2} - f(U_0) U_0^{11-m/2} }}\,,
}
where $U_0\,$ is fixed by the condition $\partial_{\theta_2}U=0\,.$
{}From now on we choose without lost of generality $U_0=U_h\,.$

{}For convenience we change variables.
In the setup \actiontau~ the $U(\theta_2)$ coordinates
are related to the cylindrical ones by
\eqn\cili{
U^{7-p} = U_h^{7-p} + U_h^{5-p} r^2\,,\quad
\theta = {2 \pi\over \delta\theta_2}\theta_2\,,
}
and these in turn with the Cartesian ones by the standard projection
$y=r \cos\theta\,,\quad z=r\sin\theta\,.$ In the latter set, the BPS
solution $U=U_h$ reads $y=0$. In the following we shall
perturb the metric around this BPS solution obtaining the spectra and
checking its stability.

Instead of finding the quark mass dependence of the
quark condensate as previously we turn to the direct evaluation of the spectra.

\subsec{Worldvolume spectra}

One of our aim is to obtain the low-energy spectrum in  the afore presented theory, similarly to \HorowitzBJ\ a discrete pattern is expected.
To work out the first order corrections to the
spectra of the worldvolume fields it suffices to expand the
action up to quadratic order. We shall deal with the correction to
the scalar/pseudoscalar and gauge fields.

\bigskip
\noindent {\it 4.1.2. Vector mesons}
\medskip

{}For gauge fields the relevant part of the Lagrangian density can be written as
\eqn\gaugelag{
{\cal L}=  -\tilde{T}_1 \alpha^{\prime 2} U^{-(14+m)/4}\left( 2 R_2^5 F_{\mu\nu}
F^{\mu\nu}
+25 {U^8\over U_h^3} F_{\mu z} F_{\nu z} \eta^{\mu\nu} \right)\,.
}
Like we look for the single states in the $S^{m-2}$
we have set to zero the component of the gauge in the compact space.
This is turn eliminates any contribution of the Chern-Simon terms.

In order to find the spectrum we expand the field contend of \gaugelag\
in a complete set of eigenfunctions in the transverse coordinate $z$:
$A_\mu(x^\mu,z) = \sum_n B_\mu^{(n)}(x^\mu)
\psi_n(z)\,, A_z(x^\mu,z) = \sum_n \varphi_\mu^{(n)}(x^\mu)
\phi_n(z)\,.$ To normalise canonically the kinetic terms of the Yang-Mills
($A_\mu$), the vectors fields ($B_\mu$) and scalar or pseudoscalar particle
($\varphi$) we use the rescaled variable $Z=z/U_h\,.$
Defining the function $K(U)=\left(U/U_h\right)^{7-p}=1+Z^2$
the gauge part of the
action derived from \gaugelag\  is proportional to
\eqn\actfs{
S_{\rm Dm} \sim -  \int dZ\, d^2x K^{-(14+m)/20} \left[
{1\over 4}  F_{\mu \nu}^{(r)} F^{\mu \nu(s)}
\psi_r \psi_s +{1\over 2} K^{8/5} m_{KK}^2
B_\mu^{(r)} B^{\mu(s)}\partial_Z \psi_r\, \partial_Z \psi_s
\right]\,,
}
with the proper normalisation condition for the $\psi$~ and $\phi$~ modes.
%\eqn\norm{
%8 \tilde{T} \alpha^{\prime 2} R_2^5 U_h^{(26-m)/4}
%\int dZ K^{(22-m)/20} \psi_r \psi_s = \delta_{rs}\,.
%}
The Born level e.o.m. for the $\psi$ mode can be derived from \actfs\
\eqn\psimode{
-K^{(14+m)/20} \partial_Z \left( K^{(18-m)/20} \partial_Z \psi_r\right)
=\lambda_r \psi_r\,,
}
where we have identify the squared mass of $\psi$ with the eigenvalue
$\lambda\,.$

%The normalisation of the scalar or pseudoscalar mode is fixed to
%\eqn\normphi{
%50 \tilde{T} \alpha^{\prime 2} U_h^{(46-m)/4}
%\int dZ K^{(54-m)/20} \phi_r \phi_s = \delta_{rs}\,.
%}
With the use of \psimode\ and the identification
$\phi_n \to m_{kk}^{-1}/\sqrt{\lambda_n} \partial_Z \psi_n\,\,(n\ge 1)\,,$
the modes with $n\ge 1$ can be gauged away by redefining $B_\mu^{(n)}\,.$
%\to B_\mu^{(n)}+\partial_\mu \varphi^{(n)}/(M_K\sqrt{\lambda_n})\,.$
This leads to the final action
\eqn\finalact{
S_{\rm Dm} = -\int d^2x \left[ {1\over 2}
\partial_\mu \varphi^{(0)} \partial^\mu \varphi^{(0)} +\sum_{n\ge 1}\left(
{1\over 4} F_{\mu\nu}^{(n)}F^{\mu\nu\,(n)}+{1\over 2} M_n^2 B_\mu^{(n)}
B^{\mu\,(n)}
\right)\right]\,.
}
The above expression indicates that the scalar/pseudoscalar and the Yang-Mills fields are massless while
the vector field $B$ acquires a mass $M_n^2=\lambda_n m_{KK}^2\,$
dictated by the eigenvalues in \psimode.

In the remainder
of this section we shall determine the massive vector spectrum. {}For that purpose
we focous in the eigenvalues of \psimode\ . We solve \psimode\ asymptotically,
at $Z\gg$. This solution together with the boundary conditions at the horizon
gives an allowed  discrete set of $\lambda_n\,.$

To search for the asymptotic form of the wave function in \psimode\  we apply the Frobenius method.  The point $z\to \infty~$ is a singular regular one. {}For  $m=4,6~$ we find two roots for the indicial equation. One
of then corresponds to a non-normalisable solution while the other leads to
$\psi \sim Z^{(m-8)/10}\,.$ {}For the case $ m=8 $ the roots of the indicial equation are degenerate and it does not contain normalisable wave functions.
In the former case the eigenfunctions can be expanded  in power series
$\psi = \sum \alpha_k Z^{-2k/5-(8-m)/10 }\,$ with coefficients
following from \psimode. Once the arbitrariness in the first one
is removed by choosing $\alpha_0=1\,,$
the asymptotic behaviour of $\psi$ in \psimode\ is fixed.
The first coefficients
are: $\alpha_1=\alpha_2=\alpha_4=\alpha_7=0\,,
\alpha_3=25\, \lambda_n/(3 m -60)\,,
\alpha_5=(144-26m+m^2)/(20m-560)\,,
\alpha_6=625 \,\lambda_n^2/(18(640-52m+m^2))\,.$

In addition to the asymptotic behaviour the regularity at the origin,
$Z=0\,,$ demands $\psi\,$ to be and even or and odd function. These
constraints pins down a discrete set of eigenvalues in \psimode\ corresponding
to the vector masses depicted in table~2.

\bigskip

{\vbox{\ninepoint{
$$
\vbox{\offinterlineskip\tabskip=0pt
\halign{\strut\vrule#
%%%%%%%%%%%%%%%%%%
%&~$#$~\hfil\vrule
%&~$#$~\hfil\vrule
&~$#$~\hfil\vrule
&~$#$~\hfil\vrule
&~$#$\hfil
&\vrule#
\cr
%%%%%%%%%%%%%%%%%%
\noalign{\hrule}
&
{\rm Eigenvalue}
&
 m=4
&
 m=6\,\,
&\cr
\noalign{\hrule}
%%%%%%%%%%%%%%%%%%
&
\quad\quad\,\,\lambda_1
&
0.309
&
0.149
&\cr
%%%%%%%%%%%%%%%%%%
&
\quad\quad\,\,\lambda_2
&
1.591
&
1.290
&\cr
%%%%%%%%%%%%%%%%%%
&
\quad\quad\,\,\lambda_3
&
3.811
&
3.349
&\cr
%%%%%%%%%%%%%%%%%%
&
\quad\quad\,\,\lambda_4
&
6.983
&
6.361
&\cr
%%%%%%%%%%%%%%%%%%
\noalign{\hrule}}
\hrule}$$
\vskip-10pt
\noindent{\bf Table 2:}
{\sl
The vector, axial-vector
 spectrum for different dimensional embbedings as obtained by
\psimode\ .
}
\vskip10pt}}}

Notice that in increasing the number of transverse directions the mass of the vector, axial-vector decreases in accordance with the absorption probabilities  for emission on the brane.

%\MinahanTM
\lref\MinahanTM{
  J.~A.~Minahan,
  ``Glueball mass spectra and other issues for supergravity duals of {QCD}
  models,''
  JHEP {\bf 9901}, 020 (1999)
  [arXiv:hep-th/9811156].
  %%CITATION = HEP-TH 9811156;%%
}

\ifig\ejdos{vector and axial-vector squared masses as a function of the state. To guide the eye we have link the states with lines. Even if in appearance seems to appears a linear behaviour this is no the case.
}{
\epsfxsize 2.7 in\epsfbox{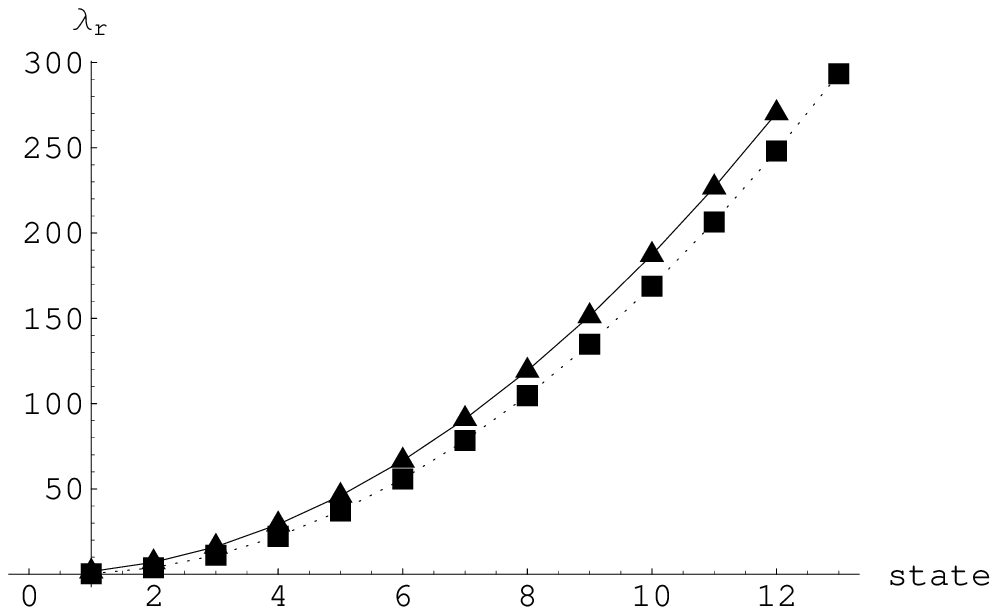}}

\smallskip
The squared masses of the vectors and vector-axial fields are represented in \ejdos. {}For both a quadratic expressions in terms of the principal quantum number, $a+b n + c n^2\,$ holds.

\bigskip
\noindent {\it 4.1.1. Massive scalar mesons}
\medskip

We are now in a position to obtain in a similar manner the massive scalar
spectrum. {}For that purpose we excite the scalar sector with the ansatz $y(x^\mu,z)\,.$ Notice that contrary to the assignment in \KruczenskiUQ\ $y$ does not depend on the $S^{m-2}$ coordinates.

%By choosing the ansatz $y(x^\mu,z)$ for the scalar field,
%the induced metric on the Dm-brane reads
%$$
%ds^2= \left( h^{-1/2} \eta_{\mu\nu} + {4\over 25} h^{1/2} \partial_\mu y
%\partial_\nu y \right) dx^\mu dx^\nu
%+{8\over 25} h^{1/2} \left( \partial_z y -  g(r) z y \right)
%\partial_\mu y\, dx^\mu dz
%$$
%\eqn\ind{
%+ {4\over 25} h^{1/2} \left( 1 - g(r) z( z  +2  y \partial_z y)
%+ (\partial_z y)^2\right) dz^2 + h^{1/2} U^2 d\Omega_{m-2}^2\,,
%}
%with $g(r) = 1/r^2 (1-(U_h/U)^{5-p})\,.$

%Defining the function $K(U)=\left(U/U_h\right)^{7-p}=1+Z^2$
In terms of the function $K\,$  the induced Dm  action  takes the form
\eqn\scalaraction{S_{\rm Dm}= -  \tilde{T}
%{4\over 25} \tilde{T} {R^5\over U_h^{(10+m)/4}}
 \int dZ d^2x \left(
{1\over 2}K^{-(14+m)/20} \partial_\mu y\, \partial^\mu y+
{1\over 2} m_{KK}^2 K^{(18-m)/20} \partial_Z y\, \partial_Z y
 +{3\over 5} m_{KK}^2  y^2
\right)\,,
}
%($\tilde{T}$ is a rescaled tension)
whose corresponding energy density is positive,
ensuring the brane stability under small perturbations.
To obtain a two-dimensional field theory description, we
expand the scalar field in a complete set of eigenfunctions in the
transverse direction:
$y(x^\mu,z)=\sum_n {\cal U}^{(n)}(x^\mu) \rho_n(z)\,.$
This allows to fix the equation of motion (e.o.m.) at  tree level for the $\rho$ mode
\eqn\scamode{
-K^{(14+m)/20} \left[\partial_Z \left( K^{(18-m)/20} \partial_Z  \rho_n \right) -\delta \rho_n\right]= \lambda_n \rho_n\,,\quad \delta={6\over 5}.
}
Notice that the previous expression is identical to \psimode\ with the only
exception of the second term in the squared brackets. Thus {\sl naively
taking the limit $\delta \to 0\,$ we shall recover the vector spectrum.}

All these ingredients together with the requirement of a canonical normalisation of the kinetic term
%With this,
%the normalisation of the kinetic term imposes the condition
%\eqn\norsca{
%- {4\over 25} \tilde{T} {R^5\over U_h^{(10+m)/4}}
 %\int dZ
%K^{-(14+m)/20} \rho_r \rho_s =\delta_{rs}\,,
%}
fixes the lagragian density
\eqn\scac{
{\cal L}= -{1\over 2}  \sum_n \left( \partial_\mu {\cal U}^{(n)}
 \partial^\mu {\cal U}^{(n)} + M_n^2 {\cal U}^{(n)}{\cal U}^{(n)} \right)\,,\quad M_n^2 =m_{KK}^2 \lambda_n\,.
 }

In order to analise the asymptotic behaviour for the different embbedings
it is convenient to perform the change of variables $Z\to \omega^{-1/a}\,,$
with $a$ been an arbitrary positive
c-number fixed only by the requirement of canceling
possible poles at $\omega\to 0\,.$ Then \scamode\ becomes
\eqn\diffdos{
\partial_\omega^2 \rho(\omega) + p(\omega) \partial_\omega \rho(\omega)
+q(\omega)\rho(\omega)=0\,,
}
with
\eqn\changesca{
\lim_{\omega\to 0} p(\omega) \to {(m+10a -8) \over 10\, a \,\omega} 
\,,\quad
\lim_{\omega\to 0}q(\omega)\to-{1\over a^2} \left(\delta\, \omega^{-(20+(2+m)/a)/10}-\lambda
\omega^{-2+6/(5a) }\right)\,,
}
where we have only displayed the divergent terms.
Notice that $\lim_{\omega \to 0} \omega\, p(\omega) \to 0$ while
$\lim_{\omega \to 0} \omega^2\, q(\omega) $ depends on the actual value of 
$a\,.$ We can choose
the coefficient $a$ to cancel either the first or the second pole in
$q(\omega)$ but not both simultaneously. In the vector case $a$ was
precisely chosen to make the $\lambda$ term in $q(\omega)$ analytic. Now
is not possible, provided $\delta \neq 0\,.$  The conclusions are far obvious:
the massive scalar wave function has not power series representation in the
asymptotic region and hence it should be a non-analytic function.
This implies that the solution of \scamode~ does not go smoothly to
the one obtained in \psimode~ as $\delta \to 0\,.$
To assess the correctness of this assertion we assume $\delta$~ to be a
free, small parameter and perform perturbation theory around the vector
solution \psimode. If correct the previous findings,
we expect to obtain the failure on the perturbation
theory assumptions at some point. We write \diffdos~ as a Schr\"odinger like
equation at zero energy with the potentials
\eqn\pot{
V(x) + \delta\,v_{\rm pert}(x) = {(m-18)\left((2+m)x^2-20\right)\over 400
(1+x^2)^2}
-{\lambda \over (1+x^2)^{8/5}} + \delta (1+x^2)^{(m-18)/20}\,.
}
The correction to the ground state energy ($E=E_0+\delta\,E_1$)~ reads
\eqn\en{
E_1= {\int_0^\infty dx \psi^2_0 (1+x^2)^{(m-18)/10}\over
\int_0^\infty dx \psi^2_0}\,,
}
where $\psi_0$~ stands for the ground state ({\sl i.e.} $\lambda_1$) wave function in
\psimode. 

The failure of perturbation theory can be traced back in the correction to the
wave function ($\rho = \psi_0 + \delta\, \psi_1$)
\eqn\wfpt{
\psi_1(x) = \psi_0(x) \int_0^x dz {1\over (1+z^2)^{(m-18)/20} \psi_0(z)^2}
\int_0^z dt \psi_0(t)^2 (1+t^2)^{(m-18)/10}\,.
}
The behaviour of $\psi_1$~ is unbounded at large distance, contrary to
that of $\psi_0\,.$ This contradict the assumption, $\lim_{x\to\infty} \psi_1
\to 0$ where \en,\wfpt~ are built in.
In conclusion the solution of \scamode~
is not an analytic function of $\delta$ at the boundary.
%This does not prevent it to have physical meaning, so far: it can well be that
%$\rho$~ is a complex function with a cut along the real axis.

%\CsakiQR
\lref\bender{
  C.~M.~ Bender, S.~A.~ Orszag,
  ``Advanced Mathematical Methods for Scientists and Engineers,''
  Springer Verlag (1991)
}

%\SternDY
\lref\SternDY{
  J.~Stern,
  ``Two alternatives of spontaneous chiral symmetry breaking in {QCD},''
  arXiv:hep-ph/9801282.
  %%CITATION = HEP-PH 9801282;%%
}

To elaborate more on the physical relevance of the solution to \scamode~
we apply the matching procedure \bender. {}For large, but finite, values
of $\lambda$~ the solutions to \scamode~ are not normalisable and hence
the model does not contain nor scalars neither pseudoscalar massive particles.
The fact that pseudoscalars particles are massless is not
in conflict with been in a chiral symmetry broken phase \SternDY:
the only necessary and sufficient criterion of
spontaneous chiral symmetry breaking (SB$\chi$S) 
is a non-zero value of the left-right
correlation function $i \lim_{m\to 0} \int\,d^4x \langle
\Omega\vert T L_\mu^i(x)R_\nu^j(0)\vert\Omega\rangle = 
-{1\over 4} \eta_{\mu\nu} \delta^{ij} F_0^2\,$ been $L_\mu$~ and $R_\mu$~ the
Noether currents generating the left and right chiral rotations respectively 
and $F_0\,,$ an order parameter,
given by $\langle 0\vert A_\mu^i\vert\pi^j \vec{p}\rangle 
= \delta^{ij} F_0 p_\mu\,.$
There are certainly many other order parameters, such as local quark 
condensates
$\langle\bar{q}q\rangle\,,\langle\bar{q}\sigma_{\mu\nu} F^{\mu\nu} q\rangle 
\ldots\,.$ A non-zero value of each of them by itself implies SB$\chi$S,
but the converse is not true {\sl i.e.} SB$\chi$S can take place ($F_0\ne 0$)
even if some of those condensate vanishes. In particular there is no 
available proof that $\langle\bar{q}q\rangle\ne 0$ is a necessary consequence of 
SB$\chi$S.

\lref\NarayananGH{
  R.~Narayanan and H.~Neuberger,
``The quark mass dependence of the pion mass at infinite N,''
  arXiv:hep-lat/0503033.
  %%CITATION = HEP-LAT 0503033;%%
}

\bigskip

In conclusion the spectrum of this second type of embedding reduces
to massless pseudoscalars and massive vectors and axials. 
To wit its reliability we compare
with the original 't Hooft model. There the full 2PI Green function amounts to
\eqn\thooft{
M_n^2 \varphi_n(x)=\left( {m_q^2-\beta^2 \over x} +
{m_{\bar{q}}^2-\beta^2 \over 1-x}\right) \varphi_n(x)-\beta^2
\int_0^1 dy\, \varphi_n(y) {\rm P} \left[{1\over(y-x)^2}\right]\,,
}
with $x\in[0,1]$ the momentum fraction carried by the quark in the light-cone
coordinates and $\beta^2 = g_{\rm strong}^2/(2 \pi) (N_c-1/N_c)\,.$ 
Furthermore $\beta$ plays the analogous role in $(1+1)d$
of $\Lambda_{\rm QCD}\,$ in $(3+1)d\,$ \GrinsteinXK .
The spectrum of \thooft\ contains a single increasing
Regge trajectory
$
M_\pi^2 \sim \beta^2 n\,, (n=1,2,\ldots)\,,
$
which is consistently found in the weak coupling regime, $
 m_q\gg g_{\rm strong}\sim {1\over\sqrt{N_c}}\,$ \ZhitnitskyUM\ and agrees
with lattice calculations \NarayananGH.
Then the lack of massive pseudoscalars in these
type of embedding discards any possibility of these been
a realistic duals to the $1+1$ 't Hooft model.

\newsec{Supergravity glueballs}

%\CsakiQR
\lref\CsakiQR{
  C.~Csaki, H.~Ooguri, Y.~Oz and J.~Terning,
  ``Glueball mass spectrum from supergravity,''
  JHEP {\bf 9901}, 017 (1999)
  [arXiv:hep-th/9806021].
  %%CITATION = HEP-TH 9806021;%%
}

%\PonsDK
\lref\PonsDK{
  J.~M.~Pons, J.~G.~Russo and P.~Talavera,
  ``Semiclassical string spectrum in a string model dual to large N QCD,''
  Nucl.\ Phys.\ B {\bf 700}, 71 (2004)
  [arXiv:hep-th/0406266].
  %%CITATION = HEP-TH 0406266;%%
}

%\PeardonJR
\lref\PeardonJR{
  M.~J.~Peardon,
  ``Coarse lattice results for glueballs and hybrids,''
  Nucl.\ Phys.\ Proc.\ Suppl.\  {\bf 63}, 22 (1998)
  [arXiv:hep-lat/9710029].
  %%CITATION = HEP-LAT 9710029;%%
}

%\DashenCK
\lref\DashenCK{
  R.~F.~Dashen, B.~Hasslacher and A.~Neveu,
  ``Nonperturbative Methods And Extended Hadron Models In Field Theory. 3.
  Four-Dimensional Nonabelian Models,''
  Phys.\ Rev.\ D {\bf 10}, 4138 (1974).
  %%CITATION = PHRVA,D10,4138;%%
}

%\AntonuccioRB
\lref\AntonuccioRB{
  F.~Antonuccio and S.~Dalley,
  ``Glueballs from (1+1)-dimensional gauge theories with transverse degrees of
  freedom,''
  Nucl.\ Phys.\ B {\bf 461}, 275 (1996)
  [arXiv:hep-ph/9506456].
  %%CITATION = HEP-PH 9506456;%%
 }

 %\BartnikAM
\lref\BartnikAM{
  R.~Bartnik and J.~Mckinnon,
  ``Particle - Like Solutions Of The Einstein Yang-Mills Equations,''
  Phys.\ Rev.\ Lett.\  {\bf 61}, 141 (1988).
  %%CITATION = PRLTA,61,141;%%
}

%\TseytlinCS
\lref\TseytlinCS{
  A.~A.~Tseytlin,
  ``On non-abelian generalisation of the Born-Infeld action in string
  theory,''
  Nucl.\ Phys.\ B {\bf 501}, 41 (1997)
  [arXiv:hep-th/9701125].
  %%CITATION = HEP-TH 9701125;%%
}

%\Gal'tsovVN
\lref\GaltsovVN{
  D.~Gal'tsov and R.~Kerner,
  ``Classical glueballs in non-Abelian Born-Infeld theory,''
  Phys.\ Rev.\ Lett.\  {\bf 84}, 5955 (2000)
  [arXiv:hep-th/9910171].
  %%CITATION = HEP-TH 9910171;%%
}

It was known long ago that classical Yang-Mills theory does not posses
in $d=2,4$ finite energy non-singular time-independent solutions \refs{\DeserWQ, \PagelsCK}.
It was also proven that neither periodic in time solutions can exist
\ColemanHD. It is then remarkable that in the AdS/CFT correspondence
a {\sl classical} background metric can account for these quantum objects at $d=3,4$ \CsakiQR.

\smallskip

In $1+1$ the situation is as follows:  in 1-dimensional space  there is no possibility of having transverse gluons. The role of these transverse gluons in the glueball formation is to fix some scale that breaks the conformal invariance of the pure Yang-Mills theory. In the absence of this scale glueballs can not exist. It is not shocking then that models with adjoint fermions \AntonuccioRB,  higgs fields \DashenCK\ or gravity \BartnikAM\ in $1+1$-dimensions contains glueballs, provided these fields couple to gluons and gives the desired scale.  The same role can be played by the use of a formalism incorporating some modification of the standard Yang-Mills density lagrangian that explicitly breaks scale invariance. This is the case of the DBI action \TseytlinCS.  This  was the precise case in the field theory side \GaltsovVN.
We shall check whether the model \metricnonex\
reflects this peculiarity.

Following \WittenZW\ we assume that the dilaton field couples to the
${\rm Tr} F^2$ operator. In Einstein
frame the e.o.m. for the scalar fluctuations are obtained via
$
\partial_\mu\left(\sqrt{g} g^{\mu \nu}\partial_\nu\tilde{\phi}
\right)=0\,.
$
In order to solve the previous expression
and calculate the mass spectrum we follow an
analogue procedure to the higher dimensional case  \CsakiQR\ and choose a
plane wave ansatz along the $\Re^{(1,1)}$ directions,
$\tilde{\phi} = e^{i  k\cdot x} \chi(U)\,.$ Using \metricnonex\ the e.o.m.
for the $d=2$~ massless dilaton boils down to
\eqn\kk{
%2 U (U^5-U_h^5) \chi''(U) + (7 U^5 + 3 U_h^5) \chi'(U) + 2 M^2 R^5
%U \chi(U) =0\,,
\partial_U\left(U^{-3/2}(U^5-U_h^5)\chi'(U)\right) + U^{-3/2} M^2 R^5 \chi(U)=0\,,\quad M^2 =-( k_1^2 - k_2^2)\,.}
%Generically denoted as $\partial_U\left(f(U)\partial_U\chi\right) +g(U)\chi=0\,.$
The asymptotic solution
depicts a normalisable term $\chi(U) \sim U^{-5/4} J_{5/6}( U^{-3/2})
 \sim U^{-5/2}$.
%\eqn\solkk{
%\chi(U)= A {M^{5/6}\over U^{5/4}} J_{-5/6}({2\over 3U^{3/2}} MR^{5/2})
%+ B {M^{5/6}\over U^{5/4}} J_{5/6}({2\over 3U^{3/2}} M R^{5/2})\,,
%}
{}For large values of $U$ \kk\ can be solved by a series solution with negative power
$\chi= \sum _{n=0}^\infty a_n U^{-(n+5/2)}U_h^{-n}\,.$ The first few non-vanishing coefficients are given by: $a_3=-2M^2R^5/(33 U_h^3)\,, a_5=1/3\,,a_6=2M^4R^{10}/(1683U_h^6)\,,$  where the arbitrary normalisation is fixed by the choice $a_0=1\,.$

The geometry must be regular at the horizon, thus we demand $\psi(U_h)$ to be regular. This fixes the possible
values of $M\,.$  We have scanned a wide range of possible values of $M$ and found no solution satisfying the proper boundary conditions.
In order to clarify the result we turn to a semi-classical evaluation of the spectra.

%\subsec{Semi-classical WKB approximation}

{\sl Semi-classical WKB approximation}:
A possible way to tackle \kk\  is to focous in highly excited states.
There the depth of the potential is
sufficient large for a semi-classical approximation be justified.
Following \MinahanTM\  we perform first a functional change
%$\psi=\left(f(U)\right)^{1/2} \chi$
after which the potential can be written as a Schr\"odinger type
equation at zero energy. %$\partial_U^2 \psi(U) - \tilde{V}(U) \psi(U)=0\,.$
Second we change
variable to  $U= F(\omega)$.
The latter step forces to choose a new multiplicative factor to bring once more the remaining expression to a Schr\"odinger type one.
%In terms of the former one the new potential reads, $
%V(\omega)=\tilde{V}(F(\omega)) (\partial_UF)^2-Q_F(\omega)/2\,,$ being $Q_F$ the Schwarzian derivative of %$F$.
All in all, applying the generic change of variables $\omega=\log(U^{7-p}-U_h^{7-p})\,$  the new potential reads (for $p=2$)
\eqn\kksc{
V(\omega)= {e^{\omega} (4 U_h^5+e^{\omega}) \over 16 (U_h^5+e^{\omega})^2}
-{M^2 R^5\over 25} {e^{\omega}\over (U_h^5+e^{\omega})^{8/5}}
\,,
}
\ifig\potkk{The potential \kksc\ . Dashed lines correspond to $d=4,3$ (short
and long dashed respectively) while full line corresponds to $d=2\,.$
We have applied to each case the change of variables
$\omega=\log(U^{7-p}-U_h^{7-p})\,.$
}{
\epsfxsize 3 in\epsfbox{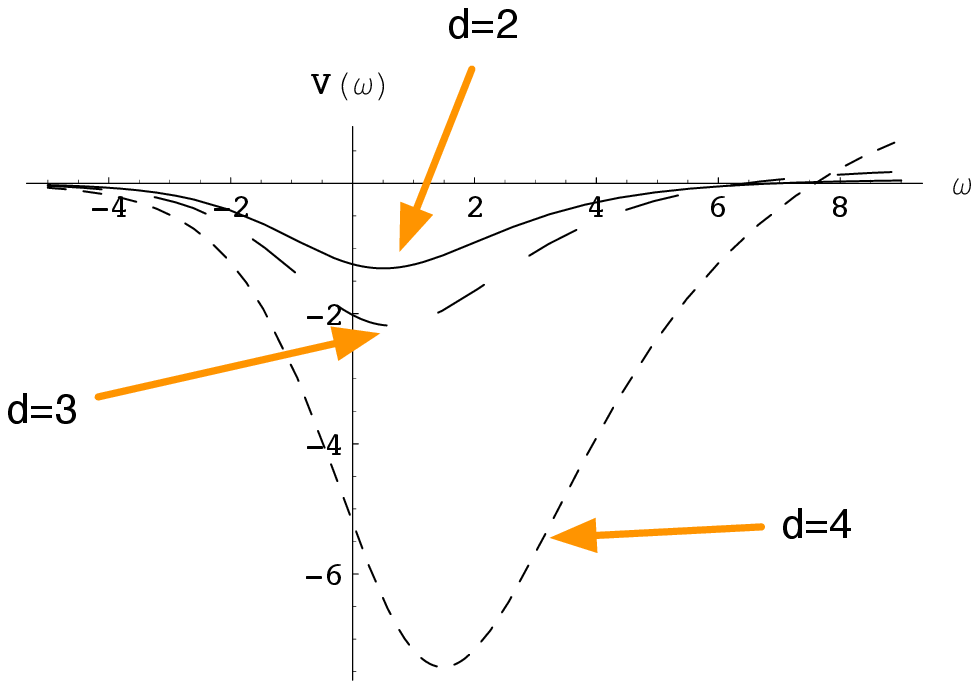}
}
that  is depicted as the full line in \potkk\ in comparison with the corresponding potentials at $d=4,3\,.$ As one can see increasing the number of transverse coordinates the depth of the well is reduced. The question is whether for $d=2$ still there is room to support at least the lowest state.

The potential in \kksc\ takes the following asymptotic forms
\eqn\potshort{
V(\omega \ll) = \left({1\over 4 U_h^5} -{M^2 R^5 \over 25 U_h^8}\right) e^{\omega} +
\left(-{7\over 16 U_h^{10}} +{8 M^2 R^5\over 125 U_h^{13}}\right) e^{2\omega}
+\cdots\,.
}
\eqn\potlong{
V(\omega \gg) = {1\over 16} - {M^2 R^5\over 25}  e^{-3 \omega/5} +
{ U_h^{5}\over 8}  e^{-\omega}
+\cdots\,.
}

Hence the classical turning point at large $\omega$ is approximately
\eqn\yplus{
\omega_+ \approx {5\over 3}\log\left({16 M^2R^5\over 25}\right)\,,
}
while the inner turning point is located at $\omega_-\to -\infty\,.$
In terms of the original variable $U$ reads
$
\omega_-=R\,.
$

Then expanding at leading order in ${1\over M^2 R^5}$ the Borh-Sommerfeld expression leads to
\eqn\wkb{
\left(n-{1\over 2}\right)\pi \approx  {1\over 5} M R^{5/2} \int_{-\infty}^{\omega_+} d\omega \,{e^{\omega/2}\over  e^\omega+U_h^5 }  +{\cal O}\left({1\over M^2 R^5}\right)\,.
}
A few remarks are in order: we discard the procedure in  \MinahanTM\ that
sets $\omega_+\to \infty\,,$ and subtract the corresponding piece as a perturbation because \wkb\ becomes divergent. This is why we should bear in mind that the upper integration limit depends on the expansion parameter.
The integration can be done analytically in terms of hypergeometric functions, but its results is not very illuminating. After this we are lead with and equality that for a given $n$ can not be saturated for a any value of $M^2\,.$

\medskip

Hence even if apparently the potential in \potkk\ has a minimum, it is not sufficient deep to hold
even the ground state. We think that this does not enter in contradiction with the results of  \GaltsovVN. First of all because in deriving \kk\ one never makes use of the DBI action and secondly because the duality must give pure YM  theory and not gluons coupled to gravity, hence the gravitons can not play the role of breaking any scale.
Although this line of reasoning, is striking that the role of the parameter $U_h$ or the coupling of the dilaton to the Liouville field in the gravity side does not provide the desired breaking, probably this is an indication that classical glueballs are protected in two dimensions also in the supergravity side.

\newsec{Breathing modes: pulsating string on $\Re^+\times \Re^+$}

In the following we shall inspect, for the general setup \metricnonex,  a point-like string configuration that corresponds to a particle
moving on the meridian of a 2-dimensional ``cigar'' shape surface, $U(\tau)\,, t(\tau)$ with the rest of coordinates constant.  In $3+1$ dimensions is was shown that this configuration was directly linked with the glueball spectra at leading order  \PonsDK. In our setup glueballs do not exist, but as we shall see a pulsating configuration still exits.
In the conformal gauge
the solutions to the e.o.m. and the Virasoro constraints
describe null geodesics of \metricnonex\
with solutions dictated by
\eqn\t{
\dot{t}={c\over 2}\left({R_2\over U}\right)^{(7-p)/2}\,,  \quad
\dot{u}= \pm {c\over 2}\sqrt{f(U)}\,.
}
This corresponds to an harmonic oscillation with target space period
\eqn\temps{
\Delta T= 2 \int_{U_h}^\infty dU \left({R_p\over U}\right)^{(7-p)/2}
{1\over \sqrt{f(U)}}\,,
}
and angular frequency $\omega_0 = 2\pi/\Delta T\,.$

To determine the spectrum we work, for convenience, with the Nambu-Goto
action considering a general setting $U(\tau)\,,t(\tau)\,, \theta_2=m\sigma\,.$
With this ansatz the lagrangian density for \metricnonex\ reduces to
$
{\cal L}= - m \left(h^{-1} f(U) - \partial_\tau{U}^2\right)^{1/2}\,,
$
that with the change of variables
$${d\zeta\over dU}= {R^{(7-p)/2}\over \sqrt{U^{7-p}-U_h^{7-p}}}$$
becomes
$ {\cal L}= - m \sqrt{h^{-1} f(\zeta)} \left(1- \partial_\tau{\zeta}^2\right)\,.
$
Its associated hamiltonian defines a one dimensional system with potential
$V=m^2 h^{-1/2} f(\zeta)/2\,.$ The semi-classical energy levels are given
by
$$
\left(2n +{1\over 2}\right) \approx
\int d\zeta  \sqrt{ E_n^2 - m^2 h^{-1/2} f(\zeta) }
$$
\eqn\levels{
=
2 \int_{U_h}^{U_1(m)}
\left({R \over U}\right)^{(7-p)/2}
 \sqrt{ E_n^2 f(U)^{-1} - m^2 \left({U\over R}\right)^{(7-p)/2} }\,,
}
where the upper turning point,
 $U_1(m)^{7-p} = (E^2/m^2) R^{(7-p)/2}-U_h^{7-p}\,,$
diverges in the massless limit. In this case and for large
energy states \levels\ reduces to
$
E_n
%\approx \omega_0 \left(n+{1\over 4}\right)
\approx n \omega_0
$
that for $p=2$~ gives
\eqn\en{
E_n\approx  - \sqrt{\pi } n \left({U_h^3\over R^5}\right)^{1/2}  {
\Gamma(-1/5) \over  \Gamma(3/10)}
\,.
}
Despite its appearance is positive defined.

%{}For two dimensions reduces to
%$\Delta T=
%-2 \sqrt{\pi }{R^{5/2}\over U_h^{3/2}}{\Gamma(3/10)\over
%\Gamma(-1/5)}\,.

%\GreensiteBE
\lref\GreensiteBE{
  J.~Greensite and M.~B.~Halpern,
  ``Suppression Of Color Screening At Large N,''
  Phys.\ Rev.\ D {\bf 27}, 2545 (1983).
  %%CITATION = PHRVA,D27,2545;%%
}

%\GrossBP
\lref\GrossBP{
  D.~J.~Gross, I.~R.~Klebanov, A.~V.~Matytsin and A.~V.~Smilga,
  ``Screening vs. Confinement in 1+1 Dimensions,''
  Nucl.\ Phys.\ B {\bf 461}, 109 (1996)
  [arXiv:hep-th/9511104].
  %%CITATION = HEP-TH 9511104;%%
}

%\ItzhakiDD
\lref\ItzhakiDD{
  N.~Itzhaki, J.~M.~Maldacena, J.~Sonnenschein and S.~Yankielowicz,
  ``Supergravity and the large N limit of theories with sixteen
  supercharges,''
  Phys.\ Rev.\ D {\bf 58}, 046004 (1998)
  [arXiv:hep-th/9802042].
  %%CITATION = HEP-TH 9802042;%%
}
\newsec{Quark-antiquark potential}

The Wilson loop is in YM the order parameter signalling confinement.
In the case of adding matter it still can be used for this purpose.
The case of YM theory coupled to adjoint matter was extensively studied in
\BrandhuberER. For the case at hand, addition of fundamental matter
supported on probe-branes, the expressions for the energy and quark-antiquark
separation are identical to those depicted in \BrandhuberER\ with the only
exception that an upper cutoff (given by the brane position in the transverse
space) appears in the integration \ParedesIS. For finite transverse distance we have
\eqn\en{
E_{\rm unsub} = {U_0\over \pi} \int_1^{U_{\rm Dm}/U_0}\,dy\,{y^{7-p}\over
\sqrt{\left(y^{7-p}-1\right)\left(y^{7-p}-\lambda\right)}}\,,
}
\eqn\long{
L= 2 {R_p^{(7-p)/2}\over U_0^{(5-p)/2}} \int_1^{U_{\rm Dm}/U_0}\,dy\,{1\over
\sqrt{\left(y^{7-p}-1\right)\left(y^{7-p}-\lambda\right)}}\,,
}
with $\lambda = (U_h/U_0)^{7-p}\,.$ In the case of dealing with probe branes at infinity one has to subtract the bare quark mass in order to obtain a finite result
\eqn\ensu{
E_{\rm sub} = E_{\rm unsub}- {U_0\over \pi} \int_1^{U_{\rm Dm}/U_0}\,dy\,\sqrt{ {y^{7-p}\over
y^{7-p}-\lambda }}\,.
}

We turn first to the asymptotic values.
The long and short string limits can be attained when the turning point
approaches
the horizon, $\lambda\to 1^+$ and $U_0\to U_{\rm  Dm}^-$ respectively. At leading order
\eqn\long{
E_{\rm long}
\approx {1\over 2\pi} \left({U_h\over R_p}\right)^{(7-p)/2} L +\ldots\,,
\quad
E_{\rm short}
\approx {1\over 2\pi}\left({U_{{\rm Dm}}\over R_p}\right)^{(7-p)/2} L + \ldots\,.
}
Thus there is an universal linear behaviour at both energy ends. We {\sl stress} that, is suggestive,  the interpretation of this phenomena at high-energy as a color screening effect: while in pure YM at large-N$_c$~ deconfinement by color screening is suppressed, \GreensiteBE, one can argue that is the addition of fundamental matter the responsible of this shape \GrossBP. 
To substantiate furthermore this claim we 
increasing the distance between the probe brane with the original stack. Then
 this linear behaviour disappears recovering the usual coulomb potential\footnote{$\dag$}{One needs to subtract then the corresponding bare quantities.}.
Eventhough we must bear in mind that we have only deal with the bosonic sector and the fermionic corrections are unavoidable for a correct treatment.
Notice also that both expressions in \long\
are afflicted with some drawbacks: increasing indiscriminately the energy we probe the compact dimension in $S^1$, while in the other extreme one can neither trust all the way to the infrared limit.
In decreasing the energy   at the point where $U\sim g_{\rm YM}^2$ the system description is in terms of $M2$ branes  \ItzhakiDD.

\medskip\

%\BrandhuberJR
\lref\BrandhuberJR{
  A.~Brandhuber and K.~Sfetsos,
  ``Wilson loops from multicentre and rotating branes, mass gaps and phase
  structure in gauge theories,''
  Adv.\ Theor.\ Math.\ Phys.\  {\bf 3}, 851 (1999)
  [arXiv:hep-th/9906201].
  %%CITATION = HEP-TH 9906201;%%
}

%\BachasXS
\lref\BachasXS{
  C.~Bachas,
  ``Convexity Of The Quarkonium Potential,''
  Phys.\ Rev.\ D {\bf 33}, 2723 (1986).
  %%CITATION = PHRVA,D33,2723;%%
}

As second and concluding remark it is instructive to look at the concavity conditions of the inter-quark
potential \BachasXS\
\eqn\concavity{
{dE_{\bar{q}q}\over dL} >0\,,\quad {d^2E_{\bar{q}q}\over dL^2} \le0\,.
}
These have already been study in several systems \BrandhuberJR.  Contrary to the systems studied
in the above reference our E vs. L relations show a smooth behaviour.
As one can verify
$\partial_{U_0} L$ is a monotonic decreasing function of $U_0$ with
zeros lying outside the physical region.
%While $\partial_{U_0} E$ displays vanishing values inside the region of interest.

\ifig\enluo{Energy as a function of the string
turning point $U_0\,$  as \en\ and \ensu. Short (long) dashed curve corresponds to $p=4,(3)\,,$
while the full curve to $p=2$
We have set $R=U_h=1\,$ for illustrative purposes.
}{
\epsfxsize 2.7 in\epsfbox{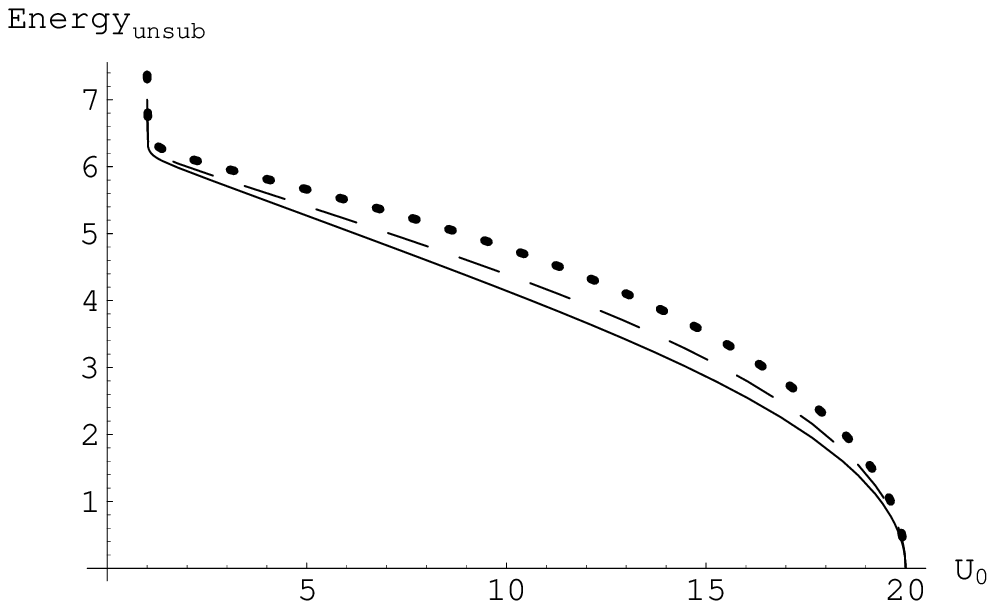}
\epsfxsize 2.7 in\epsfbox{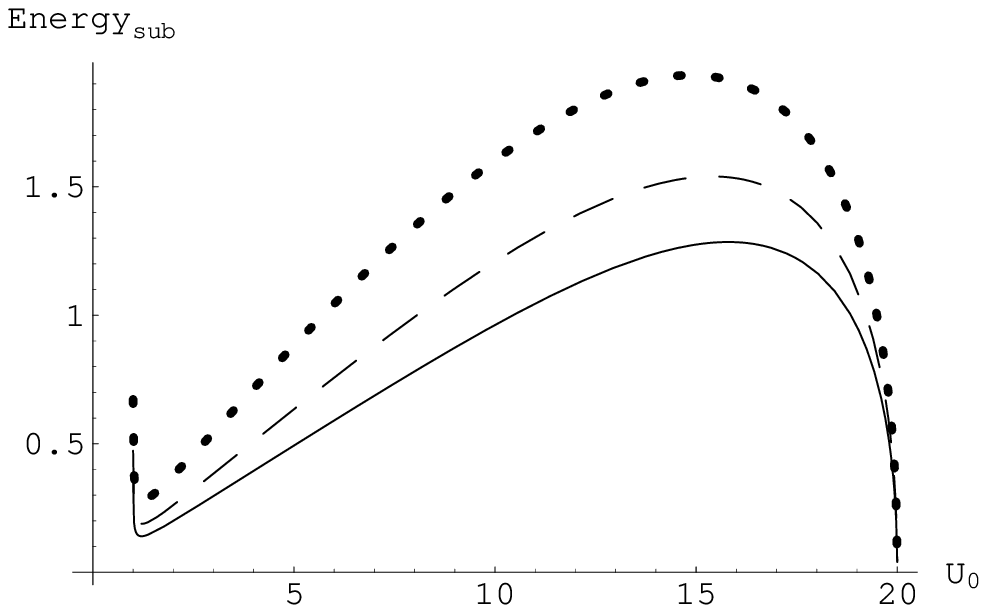}}

In \enluo\ we depict the energy versus the string turning point using a subtracted and unsubtracted relation. Notice that the expression \ensu\ has an intermediate  maximum while \en\ only takes its maximum in one of the edges. Like the difference between the two procedures is just a renormalisation scheme choice both must be equivalent, hence we conclude that with the addition of fundamental matter the shape of the subtracted energy can not be interpreted as  a phase transition phenomena.
These values of the energy must be compared with the ones needed for the opening of the compact dimensions, corresponding roughly to energies of the order of the
the Kaluza-Klein masses \kkmas\ $m_{KK}= 1.5,2,2.5$ for $p=4,3,2$ respectively. While in \en\ these suppose a constraint for the allowed region in $U_0\,,$ only short strings are allowed,  seems that \ensu\ includes also the long string case.

The expression of the energy as a function of the inter-quark separation for \en\ can be easily worked out
\eqn\ders{
{dE_{\rm unsub}\over dL} = {\lambda\over 2\pi} U_0^{(7-p)/2} + \left({E_{\rm unsub} \over U_0}
-{\lambda\over 2\pi} {U_0^{(5-p)/2}\over R_p^{(7-p)/2}} L \right) {1\over
\partial_{U_0} L}\,.
}

%\newsec{Wilson correlators}

%\newsec{Baryons}

%The volume term defines a integration constant
%\eqn\volume{
%{h^{-1}\over \sqrt{{(\partial_\sigma U)^2\over f(U)}+h^{-1}}}\,.
%}

\newsec{Summary}

We have inspected the matter content of a 1+1 dimensional field theory with the hope of finding some resemblance with QCD$_{1+1}\,$ that we know to be fully solvable in the strong coupling regime.  From the two possible general supergravity
ansatz  we have studied the first one does not lead to the correct 
relation between the quark condensate and the quark mass dependence. While for
the second type of embedding the 
spectrum reduces to massless pseudoscalars and massive vectors and axials.

Even if at first sight none of the trials to embed matter in the fundamental 
representation seems successful, 
the model contains several interesting properties as confinement and chiral 
symmetry breaking. Probably the lesson to learn is that the
most sensible way to obtain a reliable model of QCD$_{1+1}$ is not to
directly compactify on an $S^6$ but compactify QCD$_{3+1}$  or 
QCD$_{2+1}$ on an $S^2 $ or on an $S^2 $ respectively. This
would lead hopefully to the proper relation between the quark condensate
and quark mass in the first type of embeddings discussed.

Together with this, we have looked for  classical realisation of glueball in the D2-brane. Notice that this scalar sector is not the same as the one obtained in sec. 4.1.1.  The former corresponds to the Kaluza-Klein singlets states over the initial $S^6\,,$ while the latter is obtained by exciting modes on the probe brane. 
These scalars are protected to appear in accordance to the field
theory expectations. We have compared the difference with the same {\sl family}
of models but with different compact dimensions and elucidated, based on a 
semi-classical analysis, the reasons that leads to the lack of this kind of
scalars.  

{}Furthermore we have checked that even if glueballs are absent one can
obtain a cigar-like configuration. In $d=4$ the spectrum of this 
configuration reduces to that of the glueballs. This seems to be an exceptional case and we do not find deeper explanation for the coincidence.

\ifig\ejdos{The full (dashed) line describes the subtracted (unsubtracted)
energy as a function of the string length.
}{
\epsfxsize 3 in\epsfbox{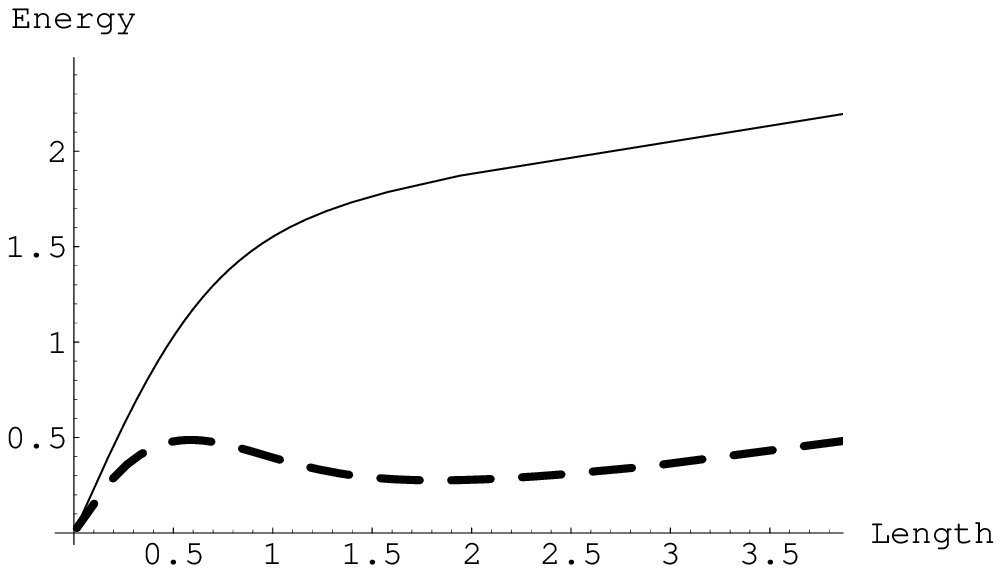}}

\smallskip
We have also looked at the quark-antiquark potential. The behaviour,
in particular the extrema,
of the energy as a function of the string turning point depends on  
a point dependent finite subtraction. The difference  
between both procedures, subtracted and unsubtracted,
 is a renormalisation scheme choice, thus
we conclude that the shapes in \ejdos~ can not be interpreted as a phase
transition.

\bigskip

{\bf Acknowledgements}

\smallskip

We thank J.~Boronat, J.~Casulleras, T.~Sakai, J.~Soto, S.~Sugimoto and O.~Varela
for discussions. Also A.~Cotrone for suggestions.
This
work is partially supported by 
the grants CYT FPA 2004-04582-C02-01, CIRIT GC 2001SGR-00065 and by
the European Community's
Human Potential Programme under contract MRTN-CT-2004-005104
`Constituents, fundamental forces and symmetries of the universe'.
M.~J.~R.~ wants to thank the Generalitat of Catalunya
for her FI-IQUC research grant.

\listrefs
\bye